\RequirePackage{lineno}
\documentclass[aps,prc,twocolumn, floatfix]{revtex4}

\usepackage{graphicx}
\usepackage{dcolumn}
\usepackage{bm}
\usepackage{color}
\usepackage{times}
\usepackage{inputenc}
\usepackage{multirow}
\usepackage{float}
\usepackage{url}
\usepackage{natbib}
\usepackage{tensor}
\usepackage{bbm}
\usepackage[colorlinks=true,citecolor=blue,urlcolor=blue,linkcolor=red]{hyperref}
\usepackage[normalem]{ulem}
\usepackage{verbatim}
\usepackage{color}

\def\Msun{\,M_{\odot}}
\def\fm3{\;\text{fm}^{-3}}

\def\mev{\;\text{MeV}}

\begin{document}

\title{Testing the phase transition parameters inside neutron stars with the production of protons and lambdas in relativistic heavy-ion collisions}

\author{Ang Li$^{1}$, Gao-Chan Yong$^{2,3}$, Ying-Xun Zhang$^{4,5}$}

\affiliation{
$^1$Department of Astronomy, Xiamen University, Xiamen, Fujian 361005, China\\
$^2$Institute of Modern Physics, Chinese Academy of Sciences, Lanzhou 730000, China\\
$^3$School of Nuclear Science and Technology, University of Chinese Academy of Sciences, Beijing 100049, China\\
$^4$China Institute of Atomic Energy, Beijing 102413, China\\
$^5$Guangxi Key Laboratory Breeding Base of Nuclear Physics and Technology, Guilin 541004, China
}

\begin{abstract}
We demonstrate the consistency of the quark deconfinement phase transition parameters in the beta-stable neutron star matter and in the nearly symmetric nuclear matter formed in heavy-ion collisions (HICs). 
We investigate the proton and $\Lambda$ flow in Au+Au collisions at 3 and 4.5 GeV/nucleon incident beam energies with the pure hadron cascade version of a multi-phase transport model.
The phase transition in HICs and neutron stars is described based on a class of hybrid equations of state from the quark mean-field model for the hadronic phase and a constant-speed-of-sound parametrization for the high-density quark phase. 
The measurements of the anisotropic proton flow at 3 GeV/nucleon by the STAR collaboration favor a relatively low phase transition density lower than $\sim 2.5$ times saturation density indicated by the gravitational wave and electromagnetic observations of neutron stars.
And the proton flow data at the higher energy of 4.5 GeV/nucleon can be used to effectively constrain the softness of high-density quark matter equations of state. 
Finally, compared to the proton flow, the $\Lambda$ flow is found to be less sensitive and not constraining to the equations of state. 

\end{abstract}

\maketitle

\section{Introduction}

It is generally believed that the degree of freedom of dense matter is hadron around nuclear saturation (number) density $n_0 \approx 0.16$~fm$^{-3}$, and quarks are liberated at asymptotically high densities. The phase structure and the equation of state [EoS; i.e., pressure-density relation $p(\varepsilon)$] of cold QCD matter for intermediate densities ($\sim1$-$10~n_0$) are unfortunately unknown. 
The EoS, however, serves as one of the important inputs not only for the evolution of heavy-ion collisions (HICs) but also in the study of neutron stars, core-collapse supernovae and binary neutron star mergers. 
Presently we focus on one of the key problems, i.e. what is the critical density for the phase transition to quark matter?
It is a matter of extensive debate not only within the theory of HICs~\cite{1996NuPhA.606..320B,2021PhLB..81536138G,2022EPJC...82..911S,2022arXiv220811996O,2019NuPhA.982..883S,2019ApJ...871..157S,2019EPJWC.20403001B}, but also in astrophysics~\cite{2018RPPh...81e6902B,2019PhRvD.100j3022H,2020ApJ...904..103M,2021PhRvC.103c5802X}.

In high-energy HICs, the two colliding nuclei produce collective motion of nucleons and fragments due to the large compression, which is named collective expands in time and reaches the surrounding detectors as baryons and mesons.
The baryon flow, which is measured, depends sensitively on the gradient of pressure developed in the fireball at the moment of maximum compression during the collision~\cite{1996NuPhA.606..320B}.
For example, flow observables at $\sqrt{s_{NN}}=2.5$-$5$ GeV are shown very sensitive to the dense nuclear matter EoS at $2$-$4n_0$~\cite{2022arXiv220811996O}.
In our previous work, the properties of phase transition of dense nuclear matter nearly symmetric nuclear matter formed in relativistic HICs is investigated~\cite{2021PhLB..81536138G}, by employing a density-dependent mean-field potential at hadronic phase and a bag-model description of quark matter.
It was shown that the results with the first-order phase transition were overall well in agreement with experimental data of proton sideward flow and proton directed flows, while those with hadronic EoSs and crossover deviated from experimental data. 

Many high-energy heavy-ion experiments have been directed toward the goal of inferring properties of the nuclear EoS~\cite{2019NuPhA.982..163G}. 
In parallel with this effort, the observations of neutron stars and binary neutron stars have long been used to infer the neutron star EoS~\cite{2007PhR...442..109L}.
Different from the HICs, in neutron stars, nuclear matter is present in beta equilibrium from very low density to several times the saturation density and is extremely neutron-rich.
The EoS of neutron star matter gives rise to a unique sequence of stellar configurations under hydrostatic equilibrium through the Tolman-Oppenheimer-Volkoff (TOV) equations and can be directly connected to the current and proposed observations of neutron stars.  

The mass measurements of massive pulsars establish a firm lower bound on the neutron star's maximum mass ($M_{\rm TOV}$). Only the EoSs that support a $M_{\rm TOV}$ larger than this lower bound can pass this constraint.
In recent years, the masses of several neutron stars are known with good precision from radio Shapiro-delay measurements~\cite{2010Natur.467.1081D,2021ApJ...915L..12F}, for example, one of  the highest mass measured is $M\sim2.07\Msun$ of PSR J0740+6620~\cite{2021ApJ...915L..12F}.
Furthermore, the improved accuracy for both masses and radii of millisecond period X-ray pulsars is reached in the measurements by NICER (Neutron Star Interior Composition Explorer)~\cite{2019ApJ...887L..21R,2019ApJ...887L..24M,2021ApJ...918L..27R,2021ApJ...918L..28M}
and the recent observation of gravitational waves (GWs) emitted during a coalescing neutron star binary has opened the door to new possibilities of obtaining information on the neutron star EoS by means of the measurement of the tidal deformability~\cite{2017PhRvL.119p1101A,2018PhRvL.121p1101A,2020ApJ...892L...3A}.
In addition, measurements are planned for the moment of inertia of PSR J0737-3039 A, the $1.338\Msun$ primary component of the first double pulsar system PSR J0737-3039, based on the long-term pulsar timing to determine the periastron advance due to relativistic spin-orbit coupling, which will provide complementary constraints on the EoS~\cite{2005MNRAS.364..635B}. 

Generally, the neutron star radius is controlled mainly by the density dependence of the nuclear symmetry energy around the nuclear saturation density $n_0$~\cite{2009AIPC.1128..131L}, while the maximum mass is a reflection of the EoS stiffness at several times the nuclear saturation density (for example, $\gtrsim5n_0$) possible in the neutron star inner core,
where a phase transition to quark matter might be present~\cite{2002PhLB..526...19B,2020ApJ...904..103M}.
Previously, incorporating the possible first-order phase transition within the QMF+CSS framework (explained later in Sec.~\ref{sec:eos})~\cite{2021ApJ...913...27L}, we obtained the neutron star EoS constrained from available multimessenger astrophysical observations of LIGO/Virgo~\cite{2017PhRvL.119p1101A,2020ApJ...892L...3A} and NICER~\cite{2019ApJ...887L..21R,2019ApJ...887L..24M,2021ApJ...918L..27R,2021ApJ...918L..28M},  
as well as the mocked data of a future moment of inertia measurement of PSR J0737-3039 A~\cite{2022MNRAS.515.5071M}, within the Bayesian statistical approach.
The analysis of the observational data is found to prefer the hadron-quark phase transition taking place at not-too-high densities ($n_{\rm trans}/n_0\approx2$) (see also in Refs.~\cite{2021PhRvC.103c5802X,2022arXiv220809085I}), and the critical transition pressure is about $16.8\mev\fm3$~\cite{2022MNRAS.515.5071M}, corresponding to a critical baryon chemical potential around $1050\mev$ for such zero-temperature beta-stable stellar matter.
Also, the sound speed squared $c_{\rm QM}^2$ above the transition should be larger than $\approx0.4$ (we work in units where $\hbar=c=1$).
Nevertheless, the strength of the phase transition (namely, how strongly the transition happens)
can not be well-determined by the neutron star global properties~\cite{2021ApJ...913...27L,2022EPJWC.26004001L}.

Since one usually assumes that there is one theoretical model of EoS that can correctly explain the nuclear matter data of different physical situations obtained in both laboratory nuclear experiments and astronomical observations~\cite{2018Univ....4...67B,2020JHEAp..28...19L,2021arXiv211208157S,2022PhRvD.105b3018T,2022Natur.606..276H}.
It is extremely interesting and helpful if the information on the nuclear EoS from HICs can be related to the physics of neutron star interiors~\cite{2022Natur.606..276H,2022arXiv220113150M,2022arXiv220901413L,2022arXiv221102224L,2022FrASS...964294G}.
The objective of the present study is to test the phase transition parameters inside neutron stars, constrained by the available multimessenger astrophysical observations, with the production of nucleons and lambdas in relativistic HICs.
For both studies of HICs and neutron stars, we employ a class of QMF + CSS EoSs, which has an explicit first-order deconfinement phase transition anticipated at finite baryon densities.
Moreover, important progress has been made recently in modelling HICs, and we utilize the so-called directed flow ($v_1$) and elliptic flow ($v_2$) in the present analysis using an extended version of relativistic transport model, namely the AMPT (A Multi-Phase Transport Model)~\cite{2005PhRvC..72f4901L}.

The paper is organized as follows. 
Section II introduces the employed  QMF + CSS framework for the EoS modelling and the AMPT for the HIC modelling, as well as the available EoS constraints for symmetric nuclear matter and neutron-rich, beta-stable stellar matter.
The results and discussions are detailed in Sec. III and the conclusion and perspective are presented in Sec. IV.

\section{Equations of state with deconfinement phase transition at symmetric and beta-stable nuclear matter} \label{sec:eos}
\begin{figure}%[tbh]%--------|---------|---------|---------|---------|---------|---------|---------|
\centering
%\vspace{-0.5cm}
\includegraphics[width=0.48\textwidth]{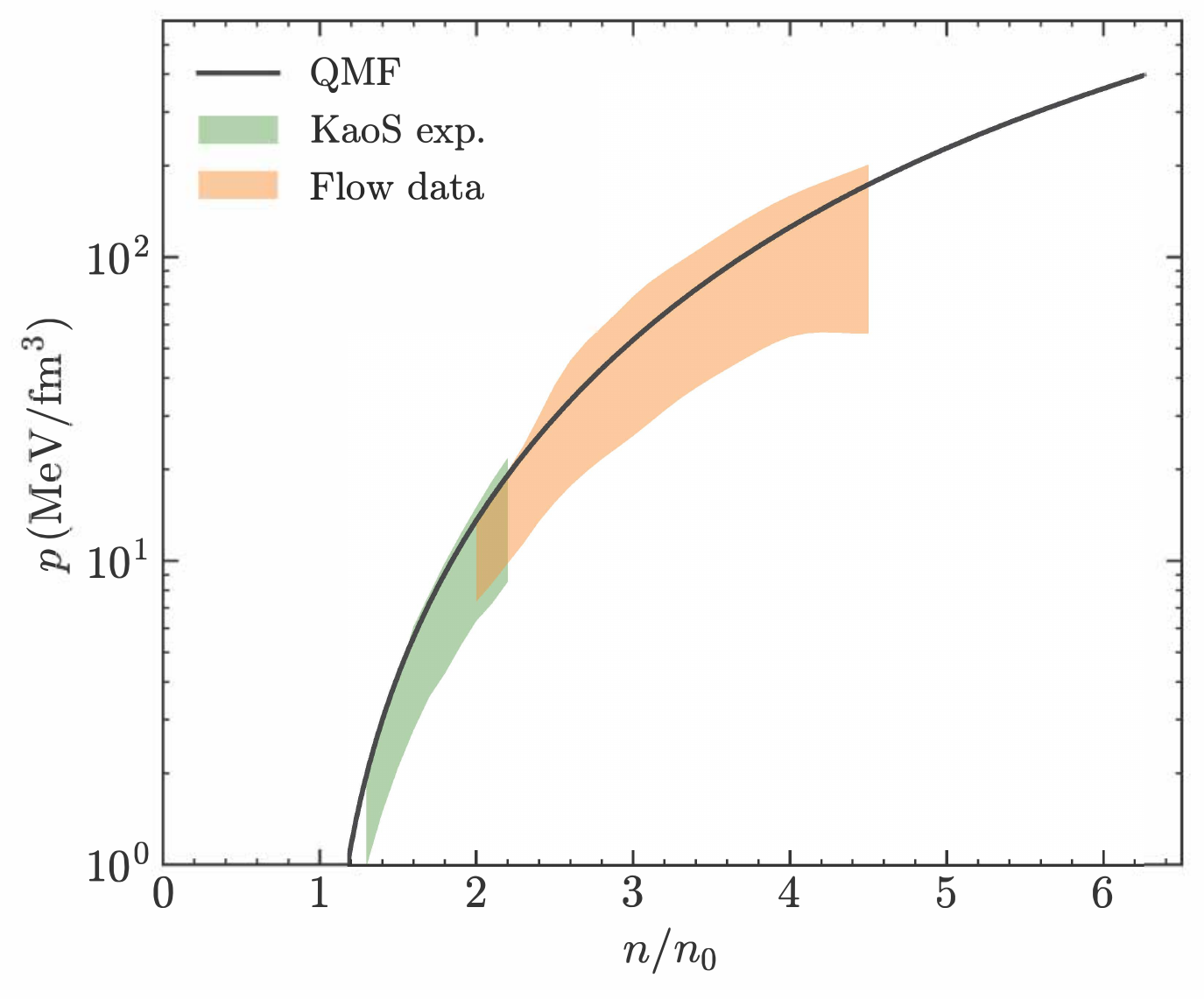} 
\caption{Pressure as a function of baryon density (in units of saturation density) of symmetric nuclear matter for QMF hadronic EoS. The shaded area at lower (higher) density corresponds to constraints~\cite{2002Sci...298.1592D} inferred from KaoS (flow) experiment. 
} \label{fig:snm}
\vspace{-0.5cm}
\end{figure}

In the present study, the EoS $p(\varepsilon)$ of nuclear matter is obtained within the quark mean-field (QMF) model (see Ref.~\cite{2020JHEAp..28...19L} for a review). 
We first adopt a harmonic oscillator potential to confine quarks in a nucleon, with its parameters determined by the mass and radius of free nucleon, and then connect the nucleon in the medium with a system of many nucleons. 
The EoS is then constructed within the widely used relativistic mean-field approach, based on an effective Lagrangian with meson fields mediating strong interactions between quarks. 
Compatibility with ab-initio calculations as well as available experimental and observational constraints at sub-nuclear and higher densities is ensured~\cite{2018ApJ...862...98Z,2019AIPC.2127b0010L}. 
For example, from the data of FOPI experiments on the elliptic flow in Au+Au collisions between 400 MeV and 1.5 GeV per nucleon, constraints can be made on the symmetric nuclear matter EoS.
Additionally, the subthreshold $K^+$ production in heavy-ion reactions measured by the Kaon Spectrometer (KaoS) collaboration has been exploited for exploring the EoS and its incompressibility.
We include both constraints~\cite{2002Sci...298.1592D} on the EoS for symmetric nuclear matter from HIC data of the KaoS experiment and flow data in Fig.~\ref{fig:snm}.
Specifically, within QMF, the saturation density is $n_0=0.16\fm3$ and the corresponding values at saturation point for the binding energy $E/A=-16\mev$, the incompressibility $K=240\mev$, the symmetry energy $E_{\rm sym}=31\mev$, the symmetry energy slope $L=40\mev$. 

%--------|---------|---------|---------|---------|---------|---------|---------|
\begin{figure}%[tbh]%--------|---------|---------|---------|---------|---------|---------|---------|
\centering
%\vspace{-0.5cm}
\includegraphics[width=0.48\textwidth]{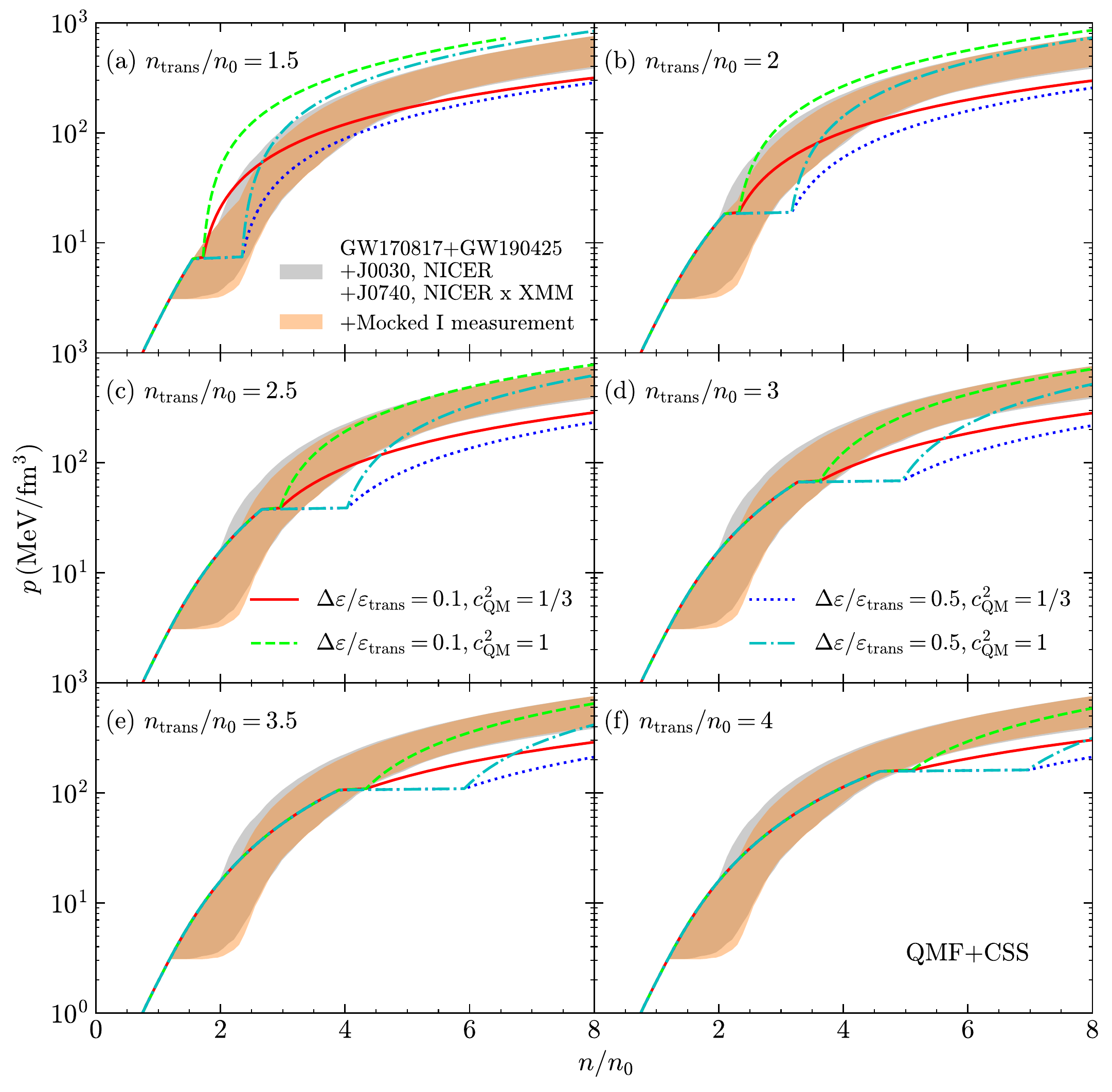}
\caption{Unified QMF hadronic EoS employed for the inner crust and outer core of neutron stars, along with various cases of CSS parameter sets: $n_{\rm trans}/n_0 = 1.5,2.0,2.5,3.0,3.5,4.0$, $\Delta\varepsilon/\varepsilon_{\rm trans}=0.1,0.5$ and $c_{\rm QM}^2=1/3,1$,
for the high-density quark matter and the hadron-quark deconfinement phase transition. 
The orange/grey shaded background are the 90\% credible regions from joint analysis~\cite{2022MNRAS.515.5071M} of the present astrophysical data, including two gravitational wave data and two NICER's mass-radius measurements, with or without a mocked moment of inertia measurement for PSR J0737-3039 A with $\sim 10\%$ accuracy possible within the next decade. 
} \label{fig:eos}
\vspace{-0.3cm}
\end{figure}%--------|---------|---------|---------|---------|---------|---------|---------|

\begin{figure}%--------|---------|---------|---------|---------|---------|---------|---------|
%\vspace{-0.3cm}
%{\centering
\includegraphics[width=0.48\textwidth]{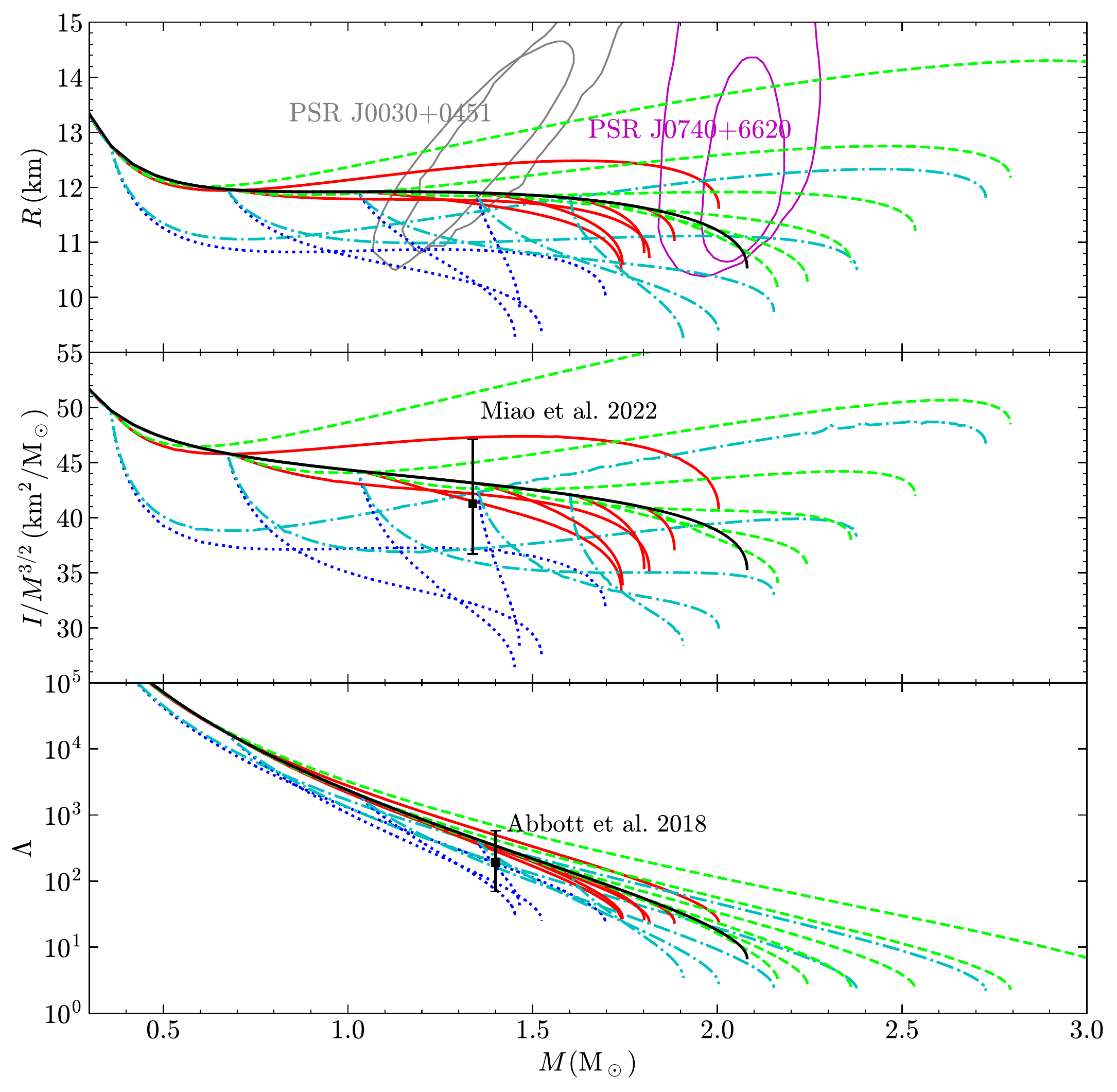}
\caption{Radius (upper panel), $I/M^{3/2}$ (middle panel), $\Lambda$ (lower panel) as functions of the stellar mass, corresponding to the 24 EoSs within QMF+CSS in Fig.~\ref{fig:eos}. 
The color coding is the same as in Fig.~\ref{fig:eos}.
And the black curves represent the results of pure hadronic stars corresponding to the QMF EoS.
In the radius-mass plot, the mass-radius measurements of the NICER telescope for PSR J0030+0451~\cite{2019ApJ...887L..24M,2019ApJ...887L..21R} 
and PSR J0740+6620~\cite{2021ApJ...918L..27R,2021ApJ...918L..28M} 
are also shown.
In the middle and lower panels, also compared are the binary tidal deformability measurement from GW170817 by LIGO/Virgo~\cite{2018PhRvL.121p1101A} and a mocked moment of inertia measurement~\cite{2022MNRAS.515.5071M} for PSR J0737-3039 A to be $I/M_{\rm A}^{3/2}=40\,{\rm km^2}/\Msun^{1/2}$ (with $\sim 10\%$ accuracy), respectively. 
}\label{fig:mr}
\vspace{-0.3cm}
\end{figure}%--------|---------|---------|---------|---------|---------|---------|---------|

For high-density quark matter and the hadron-quark deconfinement phase transition, making use of the feature that for a considerable class of microscopic quark matter models the speed of sound turns out weakly density-dependent,
we here assume the speed of sound in quark matter is density independent (see e.g., in \cite{2021ChPhC..45e5104X} for detailed discussions on the density dependence of the speed of sound in quark matter), i.e., the ``constant-speed-of-sound'' (CSS) parameterization~\cite{2013PhRvD..88h3013A} is applied.
Consequently, the EoS from the onset of the phase transition up to the maximum central density of a star is
determined by three dimensionless parameters: the transition density $n_{\rm trans}/n_0$, the transition strength $\Delta\varepsilon/\varepsilon_{\rm trans}$, and the sound speed squared in quark matter $c^2_{\rm QM}$.
The full EoS is 
\begin{equation}\label{fullEoS}
\varepsilon(p) = \left\{\!
\begin{array}{ll}
\varepsilon_{\rm HM}(p), & p<p_{\rm trans} \\ \nonumber 
\varepsilon_{\rm HM}(p_{\rm trans})+\Delta\varepsilon+c_{\rm QM}^{-2} (p-p_{\rm trans}), & p>p_{\rm trans}  
\end{array}
\right.
\end{equation}
where $n_{\rm trans} \equiv n_{\rm HM}(p_{\rm trans})$ and $\varepsilon_{\rm trans} \equiv \varepsilon_{\rm HM}(p_{\rm trans})$, and $\Delta\varepsilon/\varepsilon_{\rm trans}$ is essentially the finite discontinuity in the energy density at the phase boundary.
The high-pressure CSS EoS after the onset of the phase transition can be written as \cite{2013PhRvD..88h3013A,2020ApJ...904..103M},
\begin{eqnarray}
p(\mu_B) &=& A\mu_B^{1+1/c_{\rm QM}^2}-B\label{eq:css}\\
\mu_B (p) &=& [(p+B)/A]^{c_{\rm QM}^2/(1+c_{\rm QM}^2)}\\
n(\mu_B) &=& (1+1/c_{\rm QM}^2)A\mu_B^{1/c_{\rm QM}^2} 
\end{eqnarray}
where $A$ is a parameter with energy dimension $3-c_{\rm QM}^{-2}$ and $B=(\varepsilon_{\rm trans}+\Delta\varepsilon-c_{\rm QM}^{-2}\,p_{\rm trans})/(1+c_{\rm QM}^{-2})$.
To construct a first-order transition from some low-pressure EoS to a high-pressure EoS of Eq.~(\ref{eq:css}), $A$ is chosen such that the pressure is monotonically increasing with  $\mu_B$ and the baryon number density does not decrease at the transition. 
In practice, our study is limited to first-order phase transitions with a sharp interface (Maxwell construction). Within CSS, the matching of hadron matter to quark matter is done either in the case of beta-stable stellar matter (as shown in Fig.~\ref{fig:eos} for deducing the neutrons star observational properties; See Sec.~\ref{sec:star}) or symmetric nuclear matter (serving as input of the HIC transport simulation; See Sec.~\ref{sec:ampt}).
To avoid uncertainties due to the crust, for the computation of neutrons star properties, we also utilize unified EoSs with the consistent treatment of neutron star crusts and the crust-core transition properties along with their cores (see details in e.g., \cite{2021ApJ...913...27L}).

\subsection{Astrophysical observations of neutron star and neutron star merger}
\label{sec:star}

Neutron stars are ideal cosmic laboratories for the study of dense matter above the nuclear saturation density. 
Their global properties, such as the mass $M$, radius $R$, tidal deformability $\Lambda$ and moment of inertia $I$, have a one-to-one correspondence to the underlying EoS of neutron-rich matter in beta equilibrium. 
The calculated neutron star properties with the chosen QMF+CSS EoSs are reported in Fig.~\ref{fig:mr}, together with the available or forthcoming astrophysical constraints, respectively.
See above in Fig.~\ref{fig:eos} for the resulting regions of the EoS jointly constrained by these observational data, shown as shaded background.
For more detailed discussions on the dependence of the hybrid star mass-radius relations with the parameters of the deconfinement phase transition, we refer to Ref.~\cite{2020ApJ...904..103M}.

Among the adopted observational data, the simultaneous mass-radius measurements are for PSR J0740+6620~\cite{2021ApJ...918L..27R,2021ApJ...918L..28M} and PSR J0030+0451~\cite{2019ApJ...887L..21R,2019ApJ...887L..24M} from the NICER~collaboration. 
The dimensionless tidal deformability $\Lambda$ is related to the compactness $M/R$ and the Love number $k_2$ through $\Lambda = \frac{2}{3}k_2(M/R)^{-5}$, with the tidal Love number $k_2$ denoting the ratio of the induced quadruple moment $Q_{ij}$ to the applied tidal field $E_{ij}$,
$Q_{ij}=-k_2\frac{2R^5}{3G}E_{ij}$.
For example, the tidal deformability observations of LIGO/Virgo for the merging binary of GW170817 \cite{2017PhRvL.119p1101A} disfavor very stiff EoSs and obtained as $\Lambda(1.4)=190^{+390}_{-120}$ for a typical $1.4 M_{\odot}$ star at the $90\%$ credible level~\cite{2018PhRvL.121p1101A}.
Finally, a moment of inertia measurement of PSR J0737-3039 A, with a 90\% upper limit $I_A<3\times10^{45}\,{\rm g\ cm^2}$ is reported~\cite{2021PhRvX..11d1050K}, by fitting the post-Keplerian parameters in the binary system. 
Such a measurement gives a loose upper limit for PSR J0737-3039 A’ radius of 22 km (with 90\% confidence) and is regarded as not constraining, in comparison with the accurate observations from, e.g., LIGO/Virgo, where $R_{\rm 1.4} \leq 13.5$ km is obtained~\cite{2018PhRvL.121p1101A}.  
In the near future, the moment of inertia of PSR J0737-3039 A is expected to be measured with 11 percent precision at a 68\% confidence level in 2030~\cite{2020MNRAS.497.3118H} (indicated in Fig.~\ref{fig:mr} in the middle panel).
Both analyses with or without the forthcoming moment of inertia data are performed~\cite{2022MNRAS.515.5071M}, as reported in Fig.~\ref{fig:eos}.

\subsection{The production of nucleons and lambdas in relativistic heavy-ion collisions within AMPT}\label{sec:ampt}
Since the possible quark-gluon plasma produced in HICs is out of thermal and dynamical equilibrium~\cite{2007PrPNP..59..365W}, the study of nuclear EoS is subject to the adopted framework of transport simulations~\cite{2022arXiv220113150M}. 
We use the AMPT to model relativistic HICs, and a pure hadron cascade plus hadron mean-field potential mode AMPT-HC is used \cite{2005PhRvC..72f4901L,2021PhLB..82036521Y}. 
The AMPT model \cite{2005PhRvC..72f4901L} is known consists of four components, i.e., a fluctuating initial condition, partonic interactions, conversion from the partonic to the hadronic matter, and hadronic interactions. 
Within AMPT-HC, the dynamics of the resulting hadronic matter are described by a relativistic transport model~\cite{2001IJMPE..10..267L}.
The nucleon density distribution is provided by a Woods-Saxon parameterization and the local Thomas-Fermi approximation is employed to initialize the position and momentum of each nucleon in colliding projectile and target.
In addition to the usual elastic and inelastic collisions, hadron potentials with the test-particle method are applied to nucleons, baryon resonances, strangenesses as well as their antiparticles \cite{2021PhLB..82036521Y,2022PhRvC.106b4902Y}. 
For strange baryons $\Lambda$, $\Sigma$, $\Xi$, the quark counting rule is used \cite{1974PhRvD...9.1613M,2001PhRvL..86.2533C}.

Cross sections among various kinds of hadronic scatterings and mean-field potentials of hadrons are the main inputs of a hadronic transport model. 
In the AMPT-HC model, cross sections of hadronic scatterings can be found in Refs.~\cite{2001IJMPE..10..267L,2005PhRvC..72f4901L,2021PhLB..82036521Y} and references therein. The single baryonic potential used in the transport model is derived from the QMF EoS mentioned above. 
It is undeniable that the single nucleon potential at high energies and high densities (above $\sim2n_0$) is still unknown to date \cite{2020PhRvC.102b4913N}. The experimental Hama potential \cite{1990PhRvC..41.2737H} is effective only at the saturation density. To make minimum assumption, in this study, we use a momentum-independent single particle potential.
Specifically, $U(n)=\partial \varepsilon_{\rm pot}/\partial n$,
where the potential energy density $\varepsilon_{\rm pot}$ = $\varepsilon$ - $\varepsilon_{\rm kin}\cdot n$. 
The $\varepsilon$ is total energy density as same as in Eq.~(\ref{fullEoS}) and the nucleon average kinetic energy is usually derived from the Fermi-gas model: $\varepsilon_{\rm kin}$=${(8\pi p_{\rm F}^{5})}/({5m_{N}h^{3}n})$, with $m_{N}$ = 0.938 GeV being nucleon mass and $p_{\rm F}$ its Fermi momentum. After getting the single baryonic potential $U(n)$, the baryonic dynamics equation is expressed as $\dot{\overrightarrow{p}}$ = -$\nabla_{\overrightarrow{r}}U$. For strange baryon $\Lambda$, its potential $U_{\Lambda}(n)$ = $\frac{2}{3}U(n)$ (see e.g., \cite{2022PhRvC.106b4902Y} for more details).
We mention here that, for the matter produced in HICs, the effect of temperature dependence has been taken into account in the transport model simulations through the phase space distributions during the reactions.

\section{Results and Discussions}

Flow in general is sensitive to the EoS which governs the evolution of the system created in the nuclear collision.
At relatively higher energies, the collective flow in nuclear collisions is driven by
the pressure gradients in the early thermalized stages of the reaction and hence encodes the information about the underlying EoS, characterizing it as relatively ``hard" or  ``soft"~\cite{2001nucl.th..10037T}. 
The anisotropic flow is an important observable in characterizing how the anisotropy in the initial coordinate space develops into that in the initial momentum space, as a result of the strong interaction in the QGP matter created in relativistic HICs.

The particle directed and elliptic flow in HICs can be derived from the
Fourier expansion of the azimuthal distribution at given rapidity $y$ \cite{1994hep.ph....7282V,2008PhLB..663..312L}, i.e.,
\begin{equation}
\frac{dN}{d\phi}\propto1+2
\displaystyle{\sum_{i=1}^{n}}v_{n}\cos(n\phi) \ .
\end{equation}
The directed flow $v_{1}$ and the elliptic flow $v_{2}$ can be, respectively, expressed as:
\begin{equation}
v_{1}=\langle\cos(\phi)\rangle=\langle\frac{p_{x}}{p_{t}} \rangle \ ;%_{y}.
\end{equation}
\begin{equation}
v_{2}=\langle\cos(2\phi)\rangle=\langle\frac{p_{x}^{2}-p_{y}^{2}}{p_{t}^{2}}\rangle \ ,%_{y}.
\end{equation}
where $p_{t}$ = $\sqrt{p_{x}^{2}+p_{y}^{2}}$, and the rapidity $y$ = $\frac{1}{2}\ln\frac{E+p_{z}}{E-p_{z}}$.
\begin{figure}%[ptbh]
\centering
%\vspace{-0.3cm}
\includegraphics[width=0.48\textwidth]{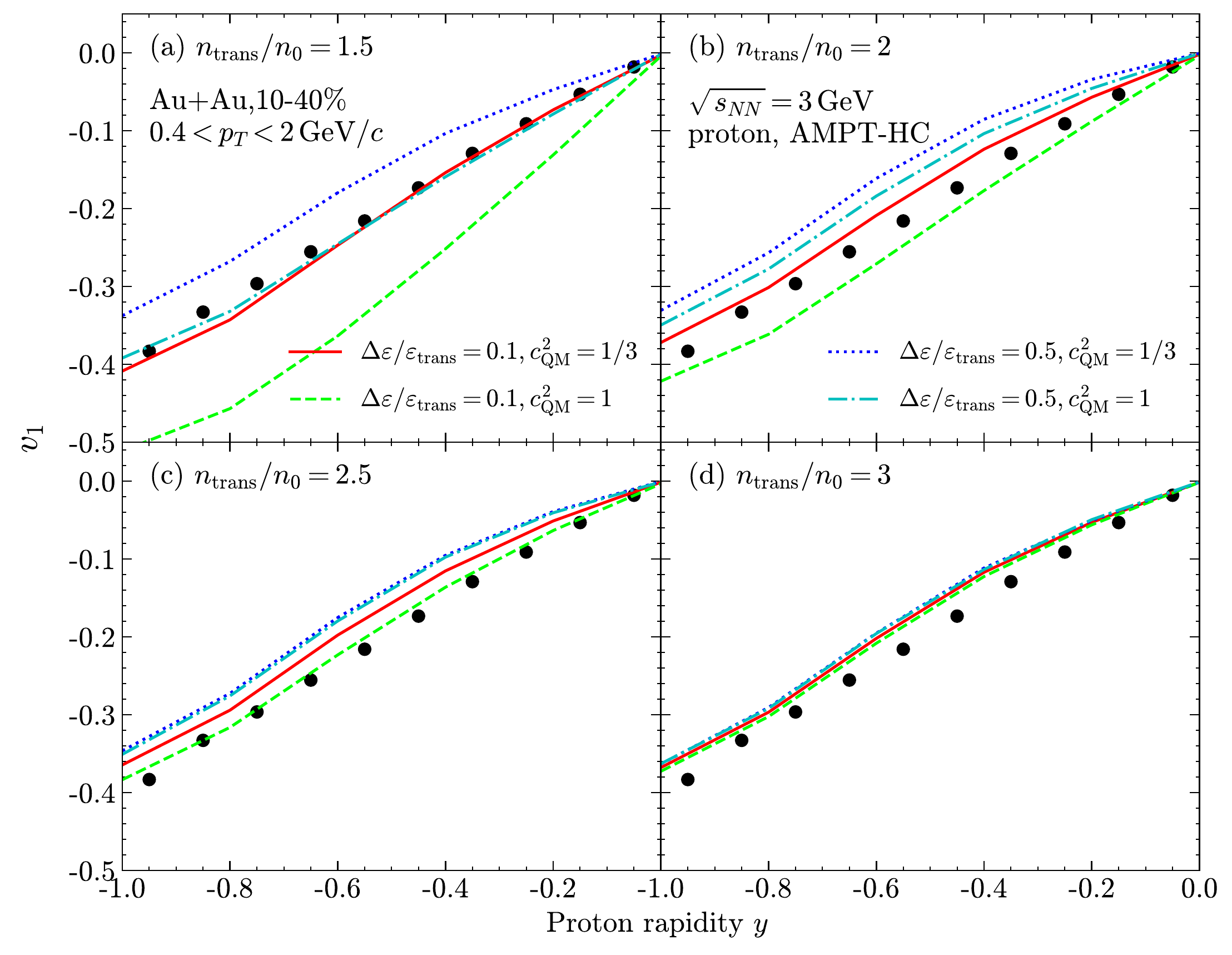}
\caption{Rapidity dependence of proton directed flow in 10-40\% centrality for Au+Au collisions at $\sqrt{s_{NN}}$ = 3 GeV by the AMPT-HC model with ($\Delta\varepsilon/\varepsilon_{\rm trans}, c_{\rm QM}^2) = (0.1, 1/3)$ (solid curves), $(0.1, 1)$ (dashed curves), $(0.5, 1/3)$ (dotted curves), $(0.5, 1)$ (dash-dotted curves), at transition density $n_{\rm trans}/n_0 = 1.5, 2.0, 2.5, 3.0$ is compared with STAR experimental data \cite{2022PhLB..82737003A}. 
} \label{v1p3g}
\vspace{-0.3cm}
\end{figure}
\begin{figure}%[ptbh]
\centering
%\vspace{-0.3cm}
\includegraphics[width=0.48\textwidth]{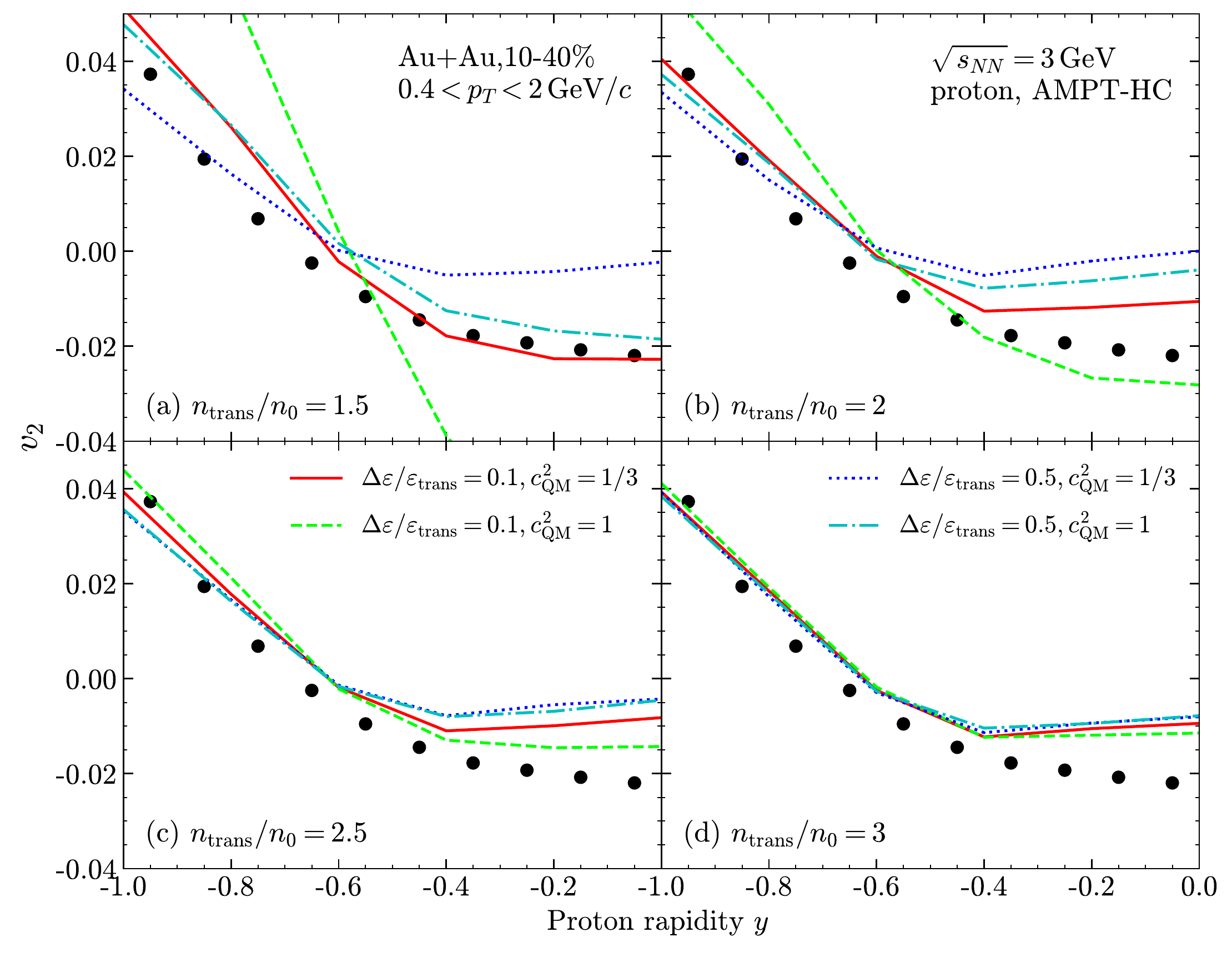}
\caption{Same with Fig.~\ref{v1p3g}, but for proton elliptic flow.
} \label{v2p3g}
\vspace{-0.3cm}
\end{figure}
\begin{figure}%[tbh]
\centering
%\vspace{-0.3cm}
\includegraphics[width=0.48\textwidth]{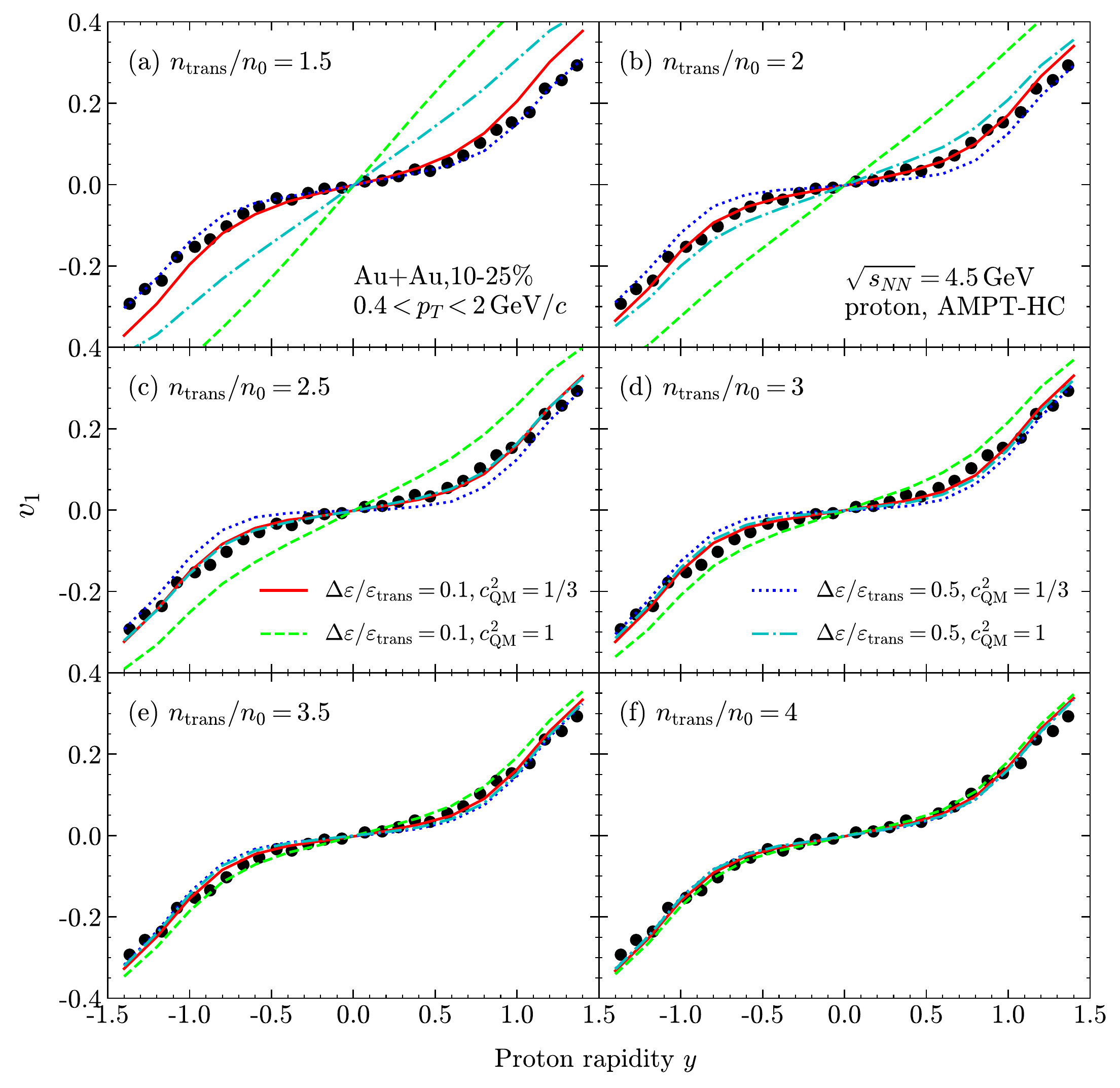}
\caption{Rapidity dependence of proton directed flow in 10-25\% centrality for Au+Au collisions at $\sqrt{s_{NN}}$ = 4.5 GeV by the AMPT-HC model with ($\Delta\varepsilon/\varepsilon_{\rm trans}, c_{\rm QM}^2) = (0.1, 1/3)$ (solid curves), $(0.1, 1)$ (dashed curves), $(0.5, 1/3)$ (dotted curves), $(0.5, 1)$ (dash-dotted curves), at transition density $n_{\rm trans}/n_0 = 1.5, 2.0, 2.5, 3.0, 3.5, 4.0$ is compared with STAR experimental data \cite{2021PhRvC.103c4908A}. 
} \label{v1p4.5g}
\vspace{-0.3cm}
\end{figure}

We first show in Fig.~4 (Fig.~5) the $v_1$ ($v_2$) of protons as a function of rapidity in 10-40\% centrality for Au+Au collisions at $\sqrt{s_{NN}}$ = 3 GeV simulated by the AMPT-HC model, for various cases of transition properties.
The full dots represent the experimental data from the STAR experiment \cite{2022PhLB..82737003A}.
The collision energy is defined as $\sqrt{s_{NN}}$ = $\sqrt{2m_{N}\times(E_{\rm lab}+2m_{N})}$ with $E_{\rm lab}$ representing the kinetic energy per nucleon of incident beam with fixed target mode.
At low transition density $n_{\rm trans}/n_0 = 1.5$ (shown in the upper-left panels of both figures), one observes an interplay between the transition strength and sound speed parameters ($\Delta\varepsilon/\varepsilon_{\rm trans},c_{\rm QM}^2$) when explaining both the $v_1$ and $v_2$ data. Neither the combination of weak first order phase transition and high speed of sound $\Delta\varepsilon/\varepsilon_{\rm trans}=0.1,c_{\rm QM}^2=1$ or strong first order phase transition and low speed of sound $\Delta\varepsilon/\varepsilon_{\rm trans}=0.5,c_{\rm QM}^2=1/3$ can explain well the experimental data. 
The case of $\Delta\varepsilon/\varepsilon_{\rm trans}=0.1,c_{\rm QM}^2=1$ for $n_{\rm trans}/n_0 = 1.5$ is especially not favored by the proton $v_2$ since the simulated $v_2$ shows no dip at central rapidities and deviate evidently from the STAR data. This is consistent with the astrophysical constraints on the phase transition in the beta-stable matter shown previously in Fig.~\ref{fig:eos} (upper-left panel), where the $(0.1,1)$ results are outside $90\%$ credible boundaries indicated by the available neutron star properties.

When comparing the results at different transition densities, it can be also seen that the experimental data points gradually shift outside of the theoretical predictions since different CSS parameters actually lead to increasingly similar results with the increase of the transition density.
Consequently, the anisotropic flow of protons at beam energy $\sqrt{s_{NN}}$ = 3 GeV can be regarded as a good probe to the transition density of possible hadron-quark phase transition and the current STAR data favor a phase transition density of $n_{\rm trans}/n_0 \lesssim 2.5$. In addition, the data seem to support a relatively modest phase transition with $\Delta\varepsilon/\varepsilon_{\rm trans}=0.1$, but tend to exclude a stiff quark matter EoS with a speed of sound as high as the speed of light ($c_{\rm QM}^2=1$).
\begin{figure}%[tbh]
\centering
%\vspace{-0.3cm}
\includegraphics[width=0.48\textwidth]{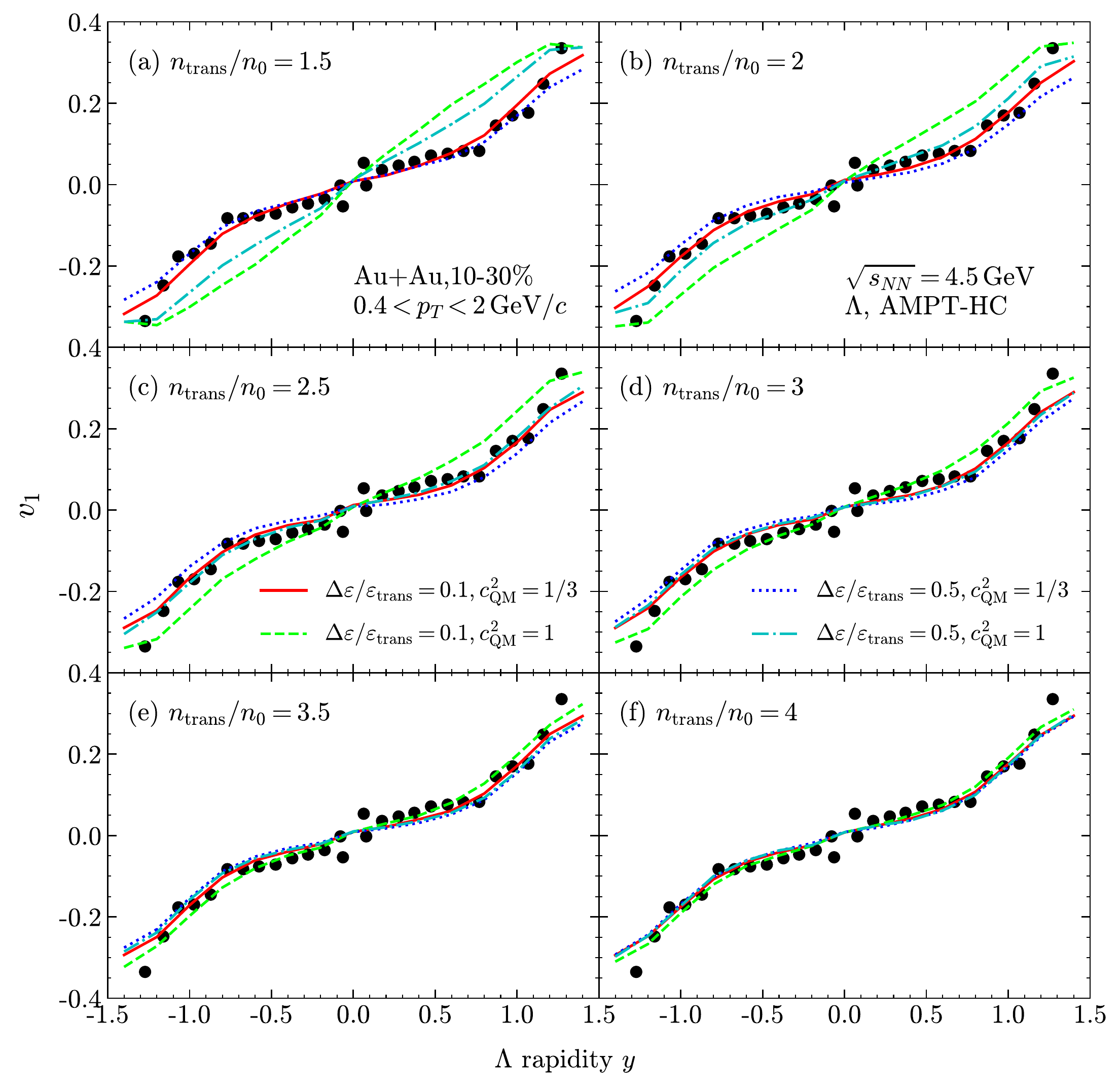}
\caption{Same with Fig.~\ref{v1p4.5g}, but for $\Lambda$ directed flow.
} \label{v1Lam4.5g}
\vspace{-0.3cm}
\end{figure}

Fig.~6 and Fig. 7 present the $v_1$ as a function of rapidity at increased beam energy $\sqrt{s_{NN}}$ = 4.5 GeV, for protons and lambdas, respectively. The full dots represent the experimental data from the STAR experiment \cite{2021PhRvC.103c4908A}.
The density in this case can reach $\gtrsim4n_0$~\cite{2021PhLB..81536138G}. 
It is seen that, for both types of hadrons, there is a change of sign of the directed flow at mid-rapidity and the $v_1$ values show in most cases the well-known bounce-off behaviour for larger rapidities, which are consistent with previous studies~\cite{1991PhR...202..233A}.
The $c_{\rm QM}^2=1$ results again fail to explain the slope of $v_1$ (no bounce-off) for low transition densities $n_{\rm trans}/n_0 \le 2.0$. Nevertheless, the flow data at $\sqrt{s_{NN}}$ = 4.5 GeV is not a good probe of the phase transition onset properties and show no sensitivity to the transition strength $\Delta\varepsilon/\varepsilon_{\rm trans}$.
Since both the proton and lambda $v_1$ STAR data support a soft quark matter close to $1/3$, the measurement at such high beam energy can be used to indicate the EoS stiffness (or softness) of high-density quark matter.
It should be noted that the lambda $v_1$ simulations are more alike under different EoSs in comparison to those of protons, therefore the lambda production is less constraining than the nucleon ones, regarding to the EoS~\cite{2022PhRvC.106d4902N}.

\section{Conclusions}
The dense matter EoS plays a major role not only in the astrophysical study related to neutron stars but also in the hydrodynamical evolution of HICs.
In this work, recent astrophysical observations of (binary)neutron stars and heavy-ion data are confronted with our present understanding of the dense QCD matter and the possible quark deconfinement phase transition. 
The advantages of using HICs to test %the constrain 
the phase transition of neutron star matter is that HICs have been operated in terrestrial laboratories for decades of years and nuclear properties have been extensively studied from different angles. 
We here make use of Au+Au collision data (directed flow and elliptic flow for both protons and lambdas) with the STAR (Solenoidal Tracker At RHIC) experiment at the beam energy $\sqrt{s_{NN}}$ = 3 and 4.5 GeV.

The dense matter in both HICs and neutron stars are described by one model of QMF+CSS, keeping consistency with available experimental and observational constraints at sub-nuclear and higher
densities. 
In our previous work, the EoS parameter space encapsulating a first-order hadron-quark phase transition has been jointly constrained by the neutron-star observational data.
We here use the obtained EoS space and the phase transition parameters (here mainly the transition density $n_{\rm trans}/n_0$, the transition strength $\Delta\varepsilon/\varepsilon_{\rm trans}$ within CSS) as a baseline for comparisons.

We find that the flow data at $\sqrt{s_{NN}}$ = 3 GeV can be used to effectively constrain the transition density, while the data at $\sqrt{s_{NN}}$ = 4.5 GeV can be used to effectively constrain the high-density quark matter EoS.
The current data support a relatively low phase transition density $\lesssim 2.5n_0$ and exclude a very stiff high-density quark matter EoS, which further supports the similar phase transition properties indicated by the astrophysical neutrons star observations.
In particular, although the transition strength can not be well signified by the neutron star properties, it might be revealed by the flow data with beam energies around $\sqrt{s_{NN}}$ = 3 GeV.
Finally, although high-energy HICs indicate a soft high-density (quark matter) EoS with the sound speed squared close to the non-interacting limit of $1/3$ (i.e., the conformal limit), there is almost no possibility to probe the possible transition from heavy-ion data with energy as high as $\sqrt{s_{NN}}=4.5$ GeV, at least based on the analysis of flow observables, due to its insensitivity of the transition threshold properties; Also, the lambda flow is found to be less sensitive than the proton flow.

There are many aspects we can explore further in the future. First, since we focus on the possible hadron-quark phase transition at the high-density nuclear matter, we here freeze the energy contribution of the isospin asymmetry part of nuclear matter following the one choice of the symmetry energy parameters at $n_0$ built in the hadronic QMF EoS. Although the choice can comfortably reconcile with various empirical constraints from low-energy nuclear physics~\cite{2014EPJA...50...37X,2021PhRvC.103a4616L}, there is still ambiguity regarding their ranges and possible intrinsic correlations between them~\cite{2020PhRvL.125t2702D}. Therefore, it is desired to incorporate the uncertainty of nuclear symmetry energy in a future study, to understand better the relevance of the phase transition taking place in symmetric nuclear matter and asymmetric matter. 
Also, more reliable constraints on EoS and phase transition parameters should be obtained in a Bayesian framework for the HIC and neutron star observational data, especially that more and higher accuracy data for both kinds are expected in the near future. 

%\section{Acknowledgments}
We are thankful to Xinle Shang, Chang Xu, Zhenyu Zhu and the XMU neutron star group for helpful discussions. The work is supported by National SKA Program of China (No.~2020SKA0120300), the National Natural Science Foundation of China (Grant Nos.~11873040,~12273028, 12275322, 11875323, 12275359) and the Youth Innovation Fund of Xiamen (No. 3502Z20206061).

\end{document}